\documentclass[runningheads]{llncs}
\usepackage[T1]{fontenc}
\usepackage{graphicx}
\begin{document}
\title{Advancing Accessibility: Augmented Reality Solutions for the Blind and Disabled in Bangladesh}
%
%
\author{Md Minhazul Islam Munna\inst{1}\ \and
Al Amin\inst{1} \and
Xin Wang\inst{1} \and
Hongbin Ma\inst{1,2*}}
\authorrunning{F. Author et al.}
%
\institute{School of Automation, Beijing Institute of Technology, Beijing
100081, People's Republic of China \\
\email{mathmhb@bit.edu.cn} \and
Laboratory{National Key Lab of Autonomous Intelligent-Unmanned Systems,
Beijing 100081, People's Republic of China}\\
}
\maketitle              
\begin{abstract}
Augmented Reality (AR) technology holds immense potential to transform the lives of blind and disabled individuals by offering enhanced interaction with their surroundings and providing real-time, accessible information. Globally, AR applications are being developed with features such as audio descriptions of environments, object recognition, and navigational aids, specifically designed to support the visually impaired. These innovations are paving the way for increased independence, mobility, and overall quality of life for millions of people worldwide. In Bangladesh, the adoption of AR technology for the blind and disabled is still in its early stages, primarily due to limited accessibility resources and infrastructure challenges. However, with the growing penetration of smartphones and continuous advancements in AR technologies, there is a promising opportunity for these solutions to be adapted, localized, and scaled within the country. This paper reviews the current state of AR technologies for the visually impaired on a global scale, explores their potential application in Bangladesh, and delves into the challenges and opportunities for broader implementation in this context.

\end{abstract}
\section{Introduction}

Augmented reality (AR) technology has shown significant potential in enhancing the quality of life for visually impaired individuals by providing real-time, context-aware assistance in navigating and interacting with their environment. Despite the availability of various assistive devices, many fail to address the specific needs of the users, limiting their daily application and effectiveness\cite{ref_article1}.
\\Recent studies have demonstrated the feasibility of using AR interfaces to improve indoor navigation for visually impaired users by integrating semantic information from the environment\cite{ref_article2}. These systems utilize a combination of visual, haptic, and auditory cues to provide a comprehensive and intuitive user experience. The growing body of research highlights the need for developing AR systems that are not only technologically advanced but also user-centered, ensuring that they cater to the unique challenges faced by the visually impaired.
\\Augmented Reality (AR) has emerged as a transformative technology with the potential to significantly enhance the lives of blind and disabled individuals. By integrating virtual elements into physical environments, AR provides accessible information and intuitive navigation aids that empower users to interact more effectively with their surroundings. Globally, AR applications are being developed to offer audio descriptions, object recognition, and real-time navigation assistance, enabling greater independence and improved quality of life for millions.
\\In Bangladesh, where resources for accessibility are limited, the adoption of AR is still in its nascent stages. However, with the increasing penetration of smartphones and advancements in AR technology, there is significant potential for these innovations to be adapted and scaled within the country. This paper reviews the current state of AR technologies for the blind and disabled globally, explores their application within Bangladesh, and discusses the challenges and opportunities for broader implementation.
\subsubsection{Motivation and Research Method: } Augmented Reality (AR) enhances accessibility for blind and disabled individuals worldwide by providing navigation assistance, object recognition, and educational tools. Applications like Microsoft’s Seeing AI offer real-time audio guidance, while AR-based rehabilitation supports therapy and social inclusion. In Bangladesh, AR can improve independence by helping visually impaired individuals navigate urban environments and access education. Additionally, AR can aid in skill development through simulated training. Successful implementation in Bangladesh will require collaboration between tech companies, the government, and NGOs, starting with pilot projects and expanding as technology becomes more accessible.
\begin{figure}
\includegraphics[width=\textwidth]{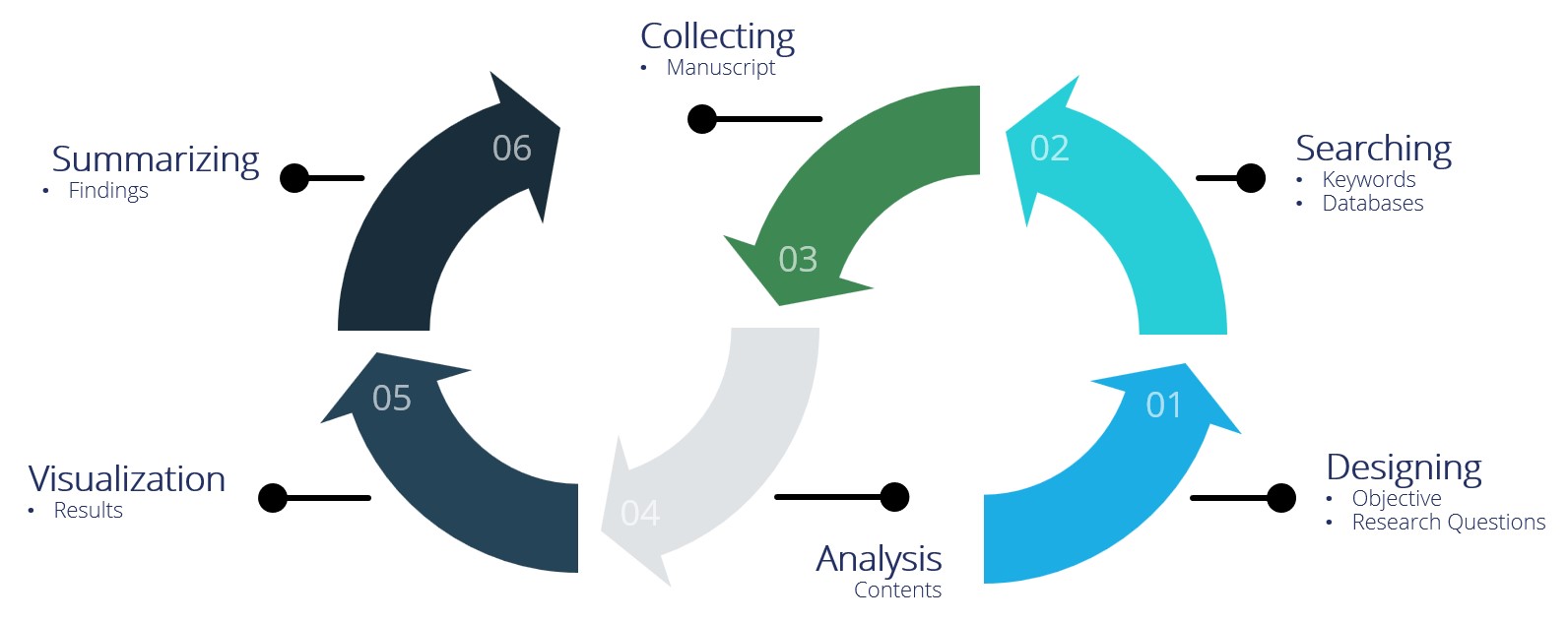}
\caption{Research Methodology} \label{fig1}
\end{figure}
Therefore, this study aims to answer the following research questions:
\\RQ1: How is Augmented Reality being used for blind and disabled individuals?
\\RQ2: What is the overall disability rate in Bangladesh?
\\RQ3: What technologies are being used in Bangladesh to assist blind and disabilities people?

The study adheres to the typical process for conducting a literature review, starting with setting objectives and formulating research questions. This is followed by searching databases, gathering relevant manuscripts, analyzing their content, visualizing the results, and summarizing the findings. Figure 1 illustrates the research design utilized throughout the study.

\section{Three Research Questions and Answers}
\subsubsection{RQ1: How is Augmented Reality being used for blind and disabled individuals?} Augmented Reality (AR) is being increasingly used to assist individuals with disabilities, including the blind, by enhancing their perception of physical information through virtual elements. For blind users, AR can provide auditory cues or haptic feedback to improve navigation and spatial awareness, incorporating features like object recognition, navigation assistance, and text-to-speech functionalities. This allows them to identify objects, read signs, and navigate unfamiliar environments with greater confidence. AR can also offer real-time language translation for individuals with communication disabilities, enabling better interaction and access to information. Additionally, AR creates interactive learning experiences tailored to the needs of individuals with disabilities, making education more engaging and accessible, thereby promoting independence, social inclusion, and participation in various activities. This contributes to a more inclusive society that values diversity and accessibility in technology and education\cite{ref_article3}.
\\Augmented Reality (AR) is being used to assist blind and disabled individuals by delivering essential knowledge in an intuitive format without relying on sensory data. Systems like CARA (Cognitive Augmented Reality Assistant) use wearable computers to capture video, extract key scene information, and convey it to users in a compact form. This aids blind users in tasks such as obstacle avoidance, scene understanding, spatial memory formation, and navigation with minimal training. AR also enables affordable virtual reality environments for testing, allowing performance comparisons, including with prosthetic devices. CARA can effectively guide users in indoor settings like public buildings and malls without physical modifications. To drive innovation in assistive technology, developers are encouraged to collaborate and create standardized test suites for evaluating and bench marking AR solutions\cite{ref_article4}.
\\Augmented Reality (AR) is used to create applications like ARTAB (Augmented Reality Tags for Assisting the Blind), designed to help blind individuals identify objects in indoor environments. ARTAB was developed using a user-centered approach, with active participation from blind users to ensure it meets their navigation and interaction needs. The application uses audio-based cues to convey the position of objects relative to the video capture device, with variations in pitch and volume indicating height and depth. This enhances spatial awareness and navigation, allowing blind users to locate objects and people effectively. Despite challenges in sound assignment, ARTAB is perceived as a valuable tool for orientation and mobility, demonstrating AR's potential to improve independence and quality of life for blind individuals by offering innovative solutions for navigation and object recognition\cite{ref_article5}.
\\Augmented Reality (AR) is being used to assist blind and disabled individuals through applications like Blind Reader, which enhances interaction with their environment. Blind Reader uses a smartphone camera to recognize objects and provides audio feedback, helping visually impaired users understand their surroundings. The app employs marker less detection, simplifying object recognition without needing specific markers. When an object is identified, the app delivers auditory feedback, aiding in decision-making and distinguishing similar objects, like different packaged foods. By integrating AR with Android smartphones, the application is widely accessible, boosting independence and confidence in daily navigation\cite{ref_article6}.
\\Augmented Reality (AR) is being used to assist blind and disabled individuals through the In Situ Audio Services (ISAS) application, which offers a practical audio AR system on smartphones. ISAS enhances orientation awareness by integrating a commercial location database with spatialized audio. Modern smartphones, equipped with GPS, compass, and sensors, enable the creation of a complete audio AR system that allows blind users to sense nearby locations like restaurants and shops, improving environmental awareness beyond simple navigation. Unlike traditional single-purpose and expensive devices, ISAS leverages widely available smartphones without additional hardware, making it more accessible and cost-effective. By utilizing a global point of interest (POI) database, ISAS offers a richer environmental experience than basic spoken text, allowing blind users to explore urban areas freely, without a specific destination. This represents a significant advancement in assistive technology, providing a practical, widely deployable solution that enhances environmental awareness and exploration\cite{ref_article7}.
\\The research paper introduces ARIANNA+, an AR-based navigation system for visually impaired individuals, enabling autonomous mobility indoors and outdoors without relying on physical landmarks like painted lines. ARIANNA+ uses the AR Kit library to create virtual paths and employs Convolutional Neural Networks (CNNs) to recognize objects and buildings, enhancing user interaction with their environment. Users navigate by loading a pre-recorded virtual path, receiving guidance through haptic, speech, and sound feedback. Portable and affordable, ARIANNA+ utilizes common smartphones, offering a cost-effective AR-based navigation solution that bridges the digital and real worlds through computer vision and machine learning, without requiring costly physical adaptations\cite{ref_article8}.
\\The research paper presents an AR interface designed to enhance environmental perception for blind users, enabling safe navigation. The system uses a haptic belt with vibrating motors controlled by a Raspberry Pi to provide tactile feedback, guiding users towards the correct travel orientation. Voice augmentation offers additional guidance, while the AR interface renders physical constraints to help users adapt and navigate safely. The system extracts landmarks like room numbers from floor maps to update navigation in real-time, aiding mental mapping of the environment. The haptic belt vibrates to guide users based on metric localization, offering real-time feedback for safe travel. The study demonstrates the feasibility of using AR to assist blind users in independent navigation, emphasizing the potential of user-oriented AR for visually impaired individuals\cite{ref_article2}.
\\Augmented Reality (AR) is used in a color recognition system to assist blind and disabled individuals by enabling effective color identification. The system features a Point-to-Sound AR algorithm that allows users to interact with colored objects using their fingers, generating sound signals to indicate colors. Users can initiate this interaction via a menu and point at objects to recognize their colors. The system displays virtual text labels of colors based on camera-captured images, with voice commands enhancing the process. Finger interaction improves user engagement, and tests showed high fingertip detection accuracy using HSV and YCbCr color spaces, proving the effectiveness of this method for color recognition in the AR system\cite{ref_article9}.
\\Augmented Reality (AR) is being explored to assist individuals with visual impairments by using augmented mirrors, which combine virtual and real content in a mirror's reflection. This approach provides an additional viewpoint for visualization and alignment, useful in both medical applications, like laparoscopic surgery, and maintenance interfaces. Virtual mirrors in AR can address challenges like missing 3D depth perception but also face occlusion issues when real-world objects block the view. The study evaluates different AR presentation modes, including in-situ rendering with X-ray techniques and side-by-side displays on CAD models or virtual mirrors, to assist with maintenance tasks in blind areas. User performance was assessed based on completion time, accuracy, and cognitive load, emphasizing the importance of user-centered AR design for enhancing accessibility and usability for individuals with disabilities. This research highlights the potential of AR technology to improve the quality of life and independence for those with visual impairments\cite{ref_article10}.
\\Augmented Reality (AR) is being harnessed to develop assistive devices that significantly improve the interaction of blind and disabled individuals with their environment. These devices offer key applications such as rapid object recognition and localization, providing auditory cues that enhance spatial awareness. By integrating binaural synthesis, AR-based auditory feedback simulates natural hearing, guiding users in navigating and interacting with objects more effectively. The use of cameras on glasses provides a stereoscopic field for 3D localization, where an auditory spatialization engine translates visual data into sound, aiding in better environmental understanding. The development of these technologies has been informed by user-centered design, focusing on practical needs like navigation in unfamiliar environments and accurate object identification. This approach ensures that the technology aligns with real-world demands, ultimately restoring cognitive abilities related to spatial awareness and navigation, thereby significantly enhancing the independence and quality of life for blind users\cite{ref_article1}.
\\Augmented Reality (AR) is being explored to assist individuals with intellectual disabilities, particularly in education and work settings. The primary focus is on enhancing work activities in horticulture by providing real-time support, helping users navigate tasks more effectively. AR enhances perception by overlaying virtual information onto the real world, making it easier for individuals with cognitive challenges to perform tasks. Wayfinding systems using AR break down complex instructions into simple steps, using images to guide users. Touch screens are highlighted as the most suitable technology for these individuals, allowing for easier navigation in mobile learning environments. However, current AR implementations can still be too complex for many users, indicating a need for further research to make these applications more accessible. Future research should focus on adhering to accessibility standards like WCAG 2.0 to ensure AR's effectiveness for people with disabilities\cite{ref_article11}.
\\Augmented Reality (AR) is being leveraged to assist blind and disabled individuals by enhancing sensory information and providing interactive experiences in real-world environments. For blind individuals, AR offers auditory cues, tactile feedback, and voice commands to aid in navigation, object recognition, and obstacle detection through audio descriptions and haptic feedback. AR can create virtual maps and navigation systems with real-time audio guidance for unfamiliar environments. For those with physical disabilities, AR supports motor skill improvement through virtual rehabilitation exercises and interactive therapy. It also aids in learning daily activities by providing visual and auditory instructions. By integrating AR into rehabilitation programs, especially for children, individuals with disabilities can engage in enjoyable activities that promote physical and cognitive development. AR also enhances social integration by facilitating communication, independent information access, and participation in educational and recreational activities. Overall, AR empowers disabled individuals to lead more independent lives, improve their quality of life, and promote inclusively\cite{ref_article12}.
\\Augmented Reality (AR) is being utilized to assist blind and disabled individuals by enhancing sensory information and providing interactive experiences. For visually impaired users, AR offers real-time audio and tactile feedback, aiding navigation, object identification, and obstacle detection. AR applications provide interactive guidance, enabling more confident exploration of new environments. The combination of AR and RFID technologies offers wheelchair users greater autonomy, particularly in Smart Cities, by facilitating independent shopping. Customized AR interfaces cater to specific needs, allowing disabled individuals to interact with physical objects and perform daily tasks with greater ease. This technology fosters inclusively by bridging the gap between the virtual and physical worlds, reducing the stigma of separation. Future advancements aim to extend AR’s benefits to a broader range of disabilities, enhancing accessibility and user experience\cite{ref_article13}.
\\Augmented Reality (AR) has been explored as a tool to provide visual supports for individuals with disabilities, including those who are blind or have neuro developmental disorders. AR can enhance user engagement by offering immersive visual aids in various activities. Research shows AR's effectiveness in teaching social skills to individuals with autism, improving eye contact, conversational engagement, and emotional recognition. Systems like Empowered Brain help direct individuals with autism towards conversational partners' faces, while AR devices like Microsoft HoloLens2 VR have shown promise in assisting with daily tasks such as making tea\cite{ref_article14}.
\\Participants in studies suggest AR's potential in teaching everyday tasks like packing a school bag or tying shoelaces, demonstrating its versatility in supporting daily living activities. AR Guides VR applications have been highlighted as valuable tools for increasing independence in children with disabilities, particularly in educational settings. However, challenges such as the affordability and accessibility of AR devices like the HoloLens2 need to be addressed to ensure wider adoption and utilization for individuals with disabilities.
\\Augmented Reality (AR) is used to assist blind and disabled individuals by providing contextualized instructions and feedback, offering multimodal sensations like visualization, animation, and contextualization. This technology benefits individuals with cognitive impairments by enhancing learning, rehabilitation, and workplace task assistance. AR increases autonomy by providing precise support typically offered by caregivers. Various AR devices—such as smart glasses, handheld devices, and projectors—are employed, with handheld devices being popular for assisted learning, especially in children, due to their affordability. Projection-based AR aids cognitively impaired adults in workplace environments, while smart glasses like the Microsoft HoloLens are gaining research attention for future use. AR enables flexible instruction designs, supporting spatial visualizations and audio, which are particularly beneficial for those with cognitive disabilities. Overall, AR has the potential to significantly enhance learning, autonomy, and task performance for blind and disabled individuals\cite{ref_article15}.

\subsubsection{RQ2: What is the overall disability rate in Bangladesh?}
The disability ratio in Bangladesh varies by gender, with 3.28\% of males and 2.32\% of females affected. The distribution of disabled individuals also differs between rural and urban areas, with 2.89\% residing in rural regions and 2.45\% in urban locales. The age distribution of disabled persons reveals that 0.83\% are children aged 0-4 years, 2.24\% are adults aged 18-49 years, and a significantly higher 9.83\% are aged 65 years and above. Regionally, Khulna division records the highest disability rate at 3.62\%, while Sylhet division has the lowest at 2.15\%. The causes of disabilities are diverse, with 41.09\% being congenital or by birth, 36.35\% resulting from illness or disease, 12.27\% from falls, such as from a tree or rooftop, and 5.53\% from road accidents\cite{ref_url16}.
\\The NSPD survey utilized a robust sampling design, collecting data from 36,000 households across the country. The classifications adhered to the Person with Disability Rights and Protection Act 2013.The primary goal of the survey is to gather data for the development and integration of persons with disabilities into the mainstream of development. It aims to ensure their equal rights in all spheres of life, focusing on employment and social inclusion.

\begin{figure}
\includegraphics[width=\textwidth]{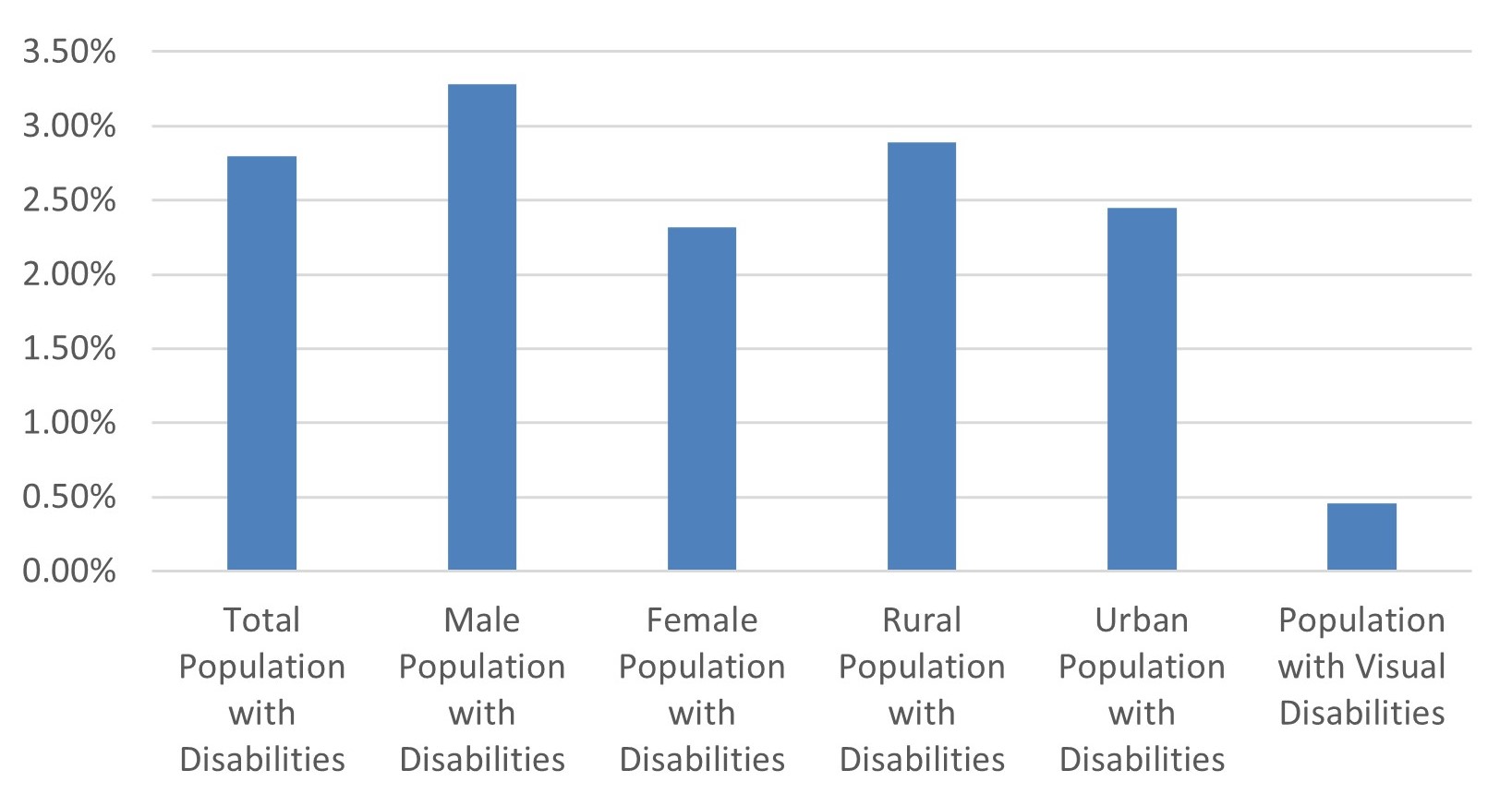}
\caption{Disability Statistics in Bangladesh (2021)} \label{fig2}
\end{figure}

In Bangladesh, over 60\% of children with disabilities are not enrolled in formal education, with only 65\% attending primary school and a mere 35\% progressing to secondary school. Those who do attend school lag academically by more than two years compared to their peers. The prevalence of disabilities among children is significant, with 1.7\% living with a disability as defined by the Persons with Disability Rights and Protection Act 2013, and 3.6\% experiencing functional difficulties in areas such as seeing, hearing, walking, and communication. Employment opportunities for persons with disabilities are limited, with only one-third of the working-age population being employed, and a marked gender disparity favoring men. Although 90\% of registered persons with disabilities receive allowances, 65\% remain unregistered, highlighting a critical gap in access to support services\cite{ref_url17}.
\\By type of disability, physical disability accounts for the highest percentage with 1.19\%, while all other types have percentages less than 1\% (autism or autism spectrum disorders: 0.04\%, mental illness leading to disability: 0.24\%, visual disability: 0.39\%, speech disability: 0.11\%, intellectual disability: 0.14\%, hearing disability: 0.19\%, cerebral palsy: 0.06\%, down syndrome: 0.03\%, deaf-blindness: 0.10\%, multiple
\\disabilities: 0.26\% and other disabilities: 0.05\%). In the case of using assistive devices by the
persons with disabilities, 18.47\% use assistive devices as per their need\cite{ref_url18}.

\subsubsection{RQ3: What technologies are being used in Bangladesh to assist blind and disabilities people?} In Bangladesh, the current technologies to assist blind students, particularly in learning mathematics, are quite limited and outdated. Blind students often use tools such as the Tailor Frame, which are not sufficient for effective learning. There is a significant demand for low-cost technological tools that can aid in learning math braille and performing calculations more easily. The study proposes a mobile phone-based interactive assistive application designed to improve learning facilities for math braille using Nemeth code for blind students in Bangladesh. This application aims to provide an effective and affordable solution by utilizing interfaces designed to meet the needs of blind students, incorporating interaction methods such as hearing and touching to facilitate self-learning\cite{ref_article19}.
\\Screen reading systems and six-point keyboards enable visually impaired (VI) individuals to access and navigate digital information efficiently. Reading machines convert printed texts into audio, while mobile audio assistance aids in transportation. Computer simulations and virtual reality create responsive environments for those with motion disabilities. ICT training programs empower people with disabilities by providing essential skills for using technology, fostering independence and societal participation\cite{ref_article20}. 
\\IoT-based smart assistants with ultrasonic sensors, along with deep learning and machine learning, improve navigation and object recognition for visually impaired users. Voice-controlled devices and wearable technologies provide audio-based assistance, enhancing accessibility and independence. GPS-enabled systems further aid in navigation, allowing visually impaired individuals to move confidently in unfamiliar environments. 
\\The primary solution available for blind individuals is external help. The paper proposes a low-cost electronic Braille display as a technological solution to assist blind people. This prototype utilizes electromagnetic solenoids controlled by an Arduino Uno to create Braille patterns, making it an affordable alternative to expensive Braille devices available internationally\cite{ref_article21}. 
\\The use of Assistive Technology (AT) in inclusive education is significant in Bangladesh, where effective AT use in educational settings is being actively studied. Robotic technology, including assistive robots, plays a crucial role in aiding visually impaired individuals with daily tasks. Notably, the Co-Robotic Cane (CRC) project focuses on developing a navigation aid to enhance independent mobility and improve the quality of life for visually impaired people. These initiatives underscore Bangladesh's commitment to leveraging technology for improving the lives of blind individuals through educational support and advanced mobility aids\cite{ref_article22}. 
\\In Bangladesh, technologies being used to assist visually impaired individuals include a smart walking stick equipped with a PIC16F690 microcontroller. This device integrates various sensors such as the Ping Sonar Sensor for obstacle detection, the GH311 Ultrasonic Obstacle Sensor for terrain mapping, and wetness detection electrodes. It provides feedback through a vibrating motor and a buzzer, offering different patterns to indicate obstacles and terrain conditions. Designed to be cost-effective and energy-efficient, the stick can be enhanced with additional features like GPS and voice guidance, making it a practical and affordable solution for visually impaired people\cite{ref_article23}. 
\\In Bangladesh, ICT tools and training for visually impaired individuals enhance educational and work opportunities, promoting social inclusion and economic development. Specialized software, Braille technology, touch screens, and talking word processors improve access to digital information and communication. Accessible web-based services further support their overall development and inclusion, playing a crucial role in expanding freedoms and opportunities\cite{ref_article24} .
\\The paper discusses the development of a smart wheelchair in Bangladesh to assist people with disabilities, featuring technologies such as voice command and joystick control, continuous remote monitoring, obstacle detection, emergency SMS alerts, a reclining back for comfort, integrated lighting for low-light conditions, and adjustable speed control to ensure safe and effective navigation\cite{ref_article25}. 
\\In Bangladesh, technologies assisting blind people include inexpensive voice-assisted smart eyewear using a Raspberry Pi 4B for processing. The eyewear employs Convolutional Neural Networks (CNN) for object classification and Optical Character Recognition (OCR) for Bangla and English text recognition, enabling users to read text audibly. Speech recognition APIs allow for voice commands to control electrical gadgets and communicate via a mobile app, enhancing interaction with the environment. Additionally, home automation systems with NodeMCU ESP32 microcontrollers enable users to manage home appliances through voice commands, significantly improving their independence and quality of life\cite{ref_article26} . 
\\In Bangladesh, various technologies assist blind people, primarily centered around mobile phones, including basic, feature, and smartphones. These devices are crucial for communication, social connection, and accessing services. Assistive applications on smartphones, such as screen readers and navigation aids, help visually impaired users perform daily tasks independently. Social media and online platforms facilitate social connections and community engagement, while community mobilizers use mobile technology to provide support, including video calls for sign language communication. Overall, these technologies enhance independence, social inclusion, and access to opportunities for visually impaired individuals in Bangladesh \cite{ref_article27}.
\\The paper highlights the use of mobile phone-based solutions, particularly the proposed mBRAILLE Android application, to assist visually impaired students (VIS) in Bangladesh with learning Braille in both Bangla and English. This approach leverages the widespread availability and affordability of mobile phones. While computer-based solutions and electronic Braille writing devices exist, they are often too expensive for most VIS in Bangladesh. The paper also mentions touch-based technologies developed in other countries, noting that many are not suitable or affordable for the local context. Overall, the focus is on creating cost-effective and accessible technological solutions to support the education of VIS in Bangladesh\cite{ref_article28}. 
\\In Bangladesh, technologies assisting blind people include widely used, affordable smartphones such as the Symphony W12 and Samsung Galaxy Star Pro, which provide a versatile platform for various applications. The Bangla Braille Learning Application (BBLA), an Android-based educational software, is specifically designed for visually impaired students to learn Braille through audio instructions and vibration feedback, minimizing the need for personal assistance. Additionally, while computers are utilized in schools, their limited availability and issues like electricity problems restrict their widespread use among students\cite{ref_article29}. 
\\the development and assessment of the Brick Blaster game, designed for visually impaired players in Bangladesh. The study reviews existing accessible games, then describes the five-level game incorporating sound, vibro-tactile, haptic feedback, and voice-over instructions. An evaluation with 24 blind participants measured usability through effectiveness, efficiency, and satisfaction. Results showed positive attitudes and high usability, with most participants successfully completing the levels, though higher levels posed more challenges. The study found no significant demographic differences in play ability, highlighting the game’s broad accessibility. The conclusion affirmed the game's usability and enjoyment, suggesting a promising approach for accessible mobile game design\cite{ref_article30} . 
\\In Bangladesh, various technologies are being employed to assist blind individuals, as demonstrated by the IoT-based blind person's stick project. The stick utilizes ultrasonic sensors for obstacle detection, with three sensors placed around the stick to alert users to obstacles on the left, right, and front through auditory messages in the user's mother tongue. An Arduino micro controller processes sensor data and controls outputs. A GSM module facilitates location tracking by sending SMS alerts to relatives in emergencies. Additionally, a speaker provides auditory feedback, and a light detection feature using an LDR sensor and LEDs signals to others that a blind person is nearby, enhancing their safety and mobility\cite{ref_article31}. 
\\In Bangladesh, technologies assisting blind people include an IoT-based smart assistant that communicates in Bengali, making it accessible for users to interact through voice commands. Additionally, a smart blind stick equipped with ultrasonic sensors, Bluetooth, and GPS/GPRS modules aids in safe navigation by detecting obstacles and providing location tracking. Furthermore, smart home integration allows blind individuals to control home appliances using Bengali voice commands, enhancing their independence and quality of life\cite{ref_article32}. 
\\These include information management systems for tracking and supporting disabled people, rehabilitation technologies inspired by Japanese practices as outlined in Bangladesh's Disability Welfare Act 2001, and NGO-driven initiatives involving approximately 70 organizations that utilize various assistive technologies. Additionally, the paper proposes systematic management of disability pensions for individuals under 20, emphasizing the role of technology in integrating disabled people into society and supporting broader developmental goals\cite{ref_article33}. 
\\In Bangladesh, various technologies are being employed to assist individuals with disabilities, significantly enhancing their quality of life. Key assistive devices include speech-generating devices that facilitate communication for those with speech impairments, exemplified by the technology used by Stephen Hawking, which can empower disabled individuals to engage in academic and social activities. Specialized software tailored for office environments is also crucial, enabling disabled users to perform tasks more efficiently. Information technology plays a vital role in gathering data about disabled individuals, which is essential for their rehabilitation and integration into society, allowing them to access necessary resources. Additionally, numerous NGOs in Bangladesh provide training and job opportunities, utilizing technology to foster skill development and create supportive environments for disabled individuals. Community-based early intervention services are also highlighted, ensuring early detection and support for disabilities, often incorporating technology to assist families in caring for disabled children. Overall, the integration of these technologies is critical for promoting inclusion and empowerment among the disabled population in Bangladesh\cite{ref_article34}. 
\\A wearable device designed to assist visually impaired individuals in Bangladesh by enabling them to identify colors. This device incorporates a TCS 3200 color sensor, which detects colors and communicates the information through an Arduino Nano micro controller. The output is delivered in both Bangla and English via connected headphones or speakers, catering to the local population's language needs. Unlike existing products such as Colorino and Speech master, which are not wearable and primarily function in English, this device emphasizes practicality and accessibility for users in Bangladesh. By enhancing color detection capabilities, the proposed technology aims to improve the daily lives of visually impaired individuals, fostering greater independence and opportunities in education and employment\cite{ref_article35}.
\\AR navigation tools offer visually impaired individuals real-time, audio-based directions through wearable devices or smartphones. By integrating GPS and object recognition technology, these tools describe surroundings, identify obstacles, and provide step-by-step guidance. For example, they can alert users when approaching a crosswalk and guide them on when to cross safely. Additionally, crowd sourced data from local users can help create a dynamic, constantly updated map of urban environments, which is especially valuable in countries like Bangladesh, where infrastructure and road conditions frequently change. AR can also enhance indoor navigation in complex environments like shopping malls, hospitals, or public buildings by using Bluetooth beacons or Wi-Fi triangulation to guide users to specific locations such as offices, restrooms, or exits\cite{ref_article36}. 

\section{Augmented Reality for the blind and disabilities people in Bangladesh}
\subsubsection{AR-Based Learning Tools} 
\subsubsection{Interactive Educational Content:} AR can revolutionize education for individuals with disabilities by creating interactive and immersive learning experiences. For example, a visually impaired student could use an AR-enabled device to feel a 3D model of the solar system through audio descriptions and haptic feedback. This would make abstract concepts more concrete and accessible.
\\\textbf{Customizable Learning Interfaces:} AR platforms could be designed to allow teachers to customize content to fit the specific needs of their students, providing personalized learning experiences. For example, a teacher could tailor a history lesson to include virtual artifacts that students can interact with, enhancing engagement and understanding.
\\\textbf{Language and Skill Training:} AR tools could offer specialized training programs for individuals with disabilities, teaching them new skills or helping them improve existing ones. For instance, a program could simulate real-world scenarios, such as job interviews or daily living tasks, to build confidence and competence.

\subsubsection{Accessible Public Services} Healthcare Accessibility: AR could be used in healthcare settings to assist both patients and healthcare providers. For instance, a blind patient could use an AR app to navigate a hospital, find the correct department, and even receive information about their appointment or medical history in an accessible format\cite{ref_article37}.
\\Public Transportation: AR applications could make public transportation more accessible for disabled individuals by providing real-time information about bus or train schedules, locations, and potential delays. The app could guide the user from their home to the nearest bus stop, notify them when their bus is approaching, and even help them find an available seat\cite{ref_article38}.
\\Government Services: AR could improve access to government services by making forms, procedures, and information more accessible. For example, an AR app could guide a visually impaired person through the process of renewing their ID card, ensuring that they complete each step correctly\cite{ref_article39}.

\subsubsection{Challenges} One significant barrier is the lack of widespread technological infrastructure, which could hinder the effective implementation of AR solutions. Additionally, the cost of deploying AR technologies might be prohibitive, requiring substantial investment and support from the government or NGOs. Furthermore, there is a need for greater awareness among the population about the potential benefits of AR, particularly for marginalized communities, to ensure its successful integration and utilization.
\\To make AR for the blind and physically challenged a reality in Bangladesh, collaboration between tech companies, government bodies, and non-governmental organizations will be crucial. Starting with pilot projects and gradually scaling up as the technology becomes more affordable and accessible could be a viable strategy\cite{ref_article40}.

\subsubsection{Implementation Strategy}
\subsubsection{Pilot Projects:} To begin, small-scale pilot projects could be launched in urban areas with high foot traffic, such as Dhaka. These pilots would help identify potential challenges and refine the technology before wider deployment.
\\\textbf{Collaboration:} Successful implementation will require collaboration between tech companies, NGOs, government bodies, and local communities. Partnerships could help secure funding, raise awareness, and ensure that the technology is accessible to those who need it most.
\\\textbf{Training and Support:} Users will need training to effectively use AR tools, especially those who are not familiar with modern technology. Offering workshops and support services could help overcome this barrier.
The opportunities for AR in Bangladesh are vast, especially when tailored to the needs of the blind and physically challenged. By focusing on navigation, education, and public service accessibility, AR can significantly improve the quality of life for these individuals, empowering them to lead more independent and fulfilling lives. With careful planning and collaboration, these technologies could be successfully integrated into Bangladesh’s societal framework.

\section{Conclusions} The paper provides an extensive review of the use of Augmented Reality (AR) technologies to assist blind and disabled individuals globally, with a specific focus on the context of Bangladesh. AR has the potential to significantly enhance the lives of these individuals by providing innovative solutions for navigation, object recognition, and environmental interaction. Globally, AR technologies are being used to create audio-visual aids that assist in spatial awareness, obstacle detection, and navigation, among other tasks, which can substantially improve the quality of life and independence for users.
\\However, in Bangladesh, the adoption of such technologies is still in its early stages. Limited resources, outdated tools, and a lack of awareness are significant barriers to the widespread implementation of AR solutions. Despite these challenges, there is considerable potential for the integration of AR in assisting the blind and disabled in Bangladesh, particularly with the growing availability of smartphones and the advancement of AR applications. The paper suggests that with proper adaptation and scaling, AR technologies could be a cost-effective and impactful solution in this context.
\\The review highlights the importance of user-centered design, affordability, and the need for collaboration among developers to create standardized solutions. It also emphasizes the necessity for further research and development to make AR technologies more accessible and practical for the disabled population in Bangladesh. Ultimately, the paper concludes that while there are significant challenges to overcome, the potential benefits of AR for the disabled are immense, and efforts should be made to integrate these technologies into the mainstream to promote greater independence and inclusion for all.

%
%
%

\begin{thebibliography}{8}
\bibitem{ref_article5}
Sánchez J, Tadres A. Augmented reality application for the navigation of people who are blind.
\end{thebibliography}
%

\end{document}